\documentclass[aps,pre,preprint,floatfix]{revtex4}
\usepackage{graphicx}

\begin{document}
\title{ Heat transport in turbulent Rayleigh-B\'enard convection: Effect of finite top- and bottom-plate conductivity} 
\author{Eric Brown}
\author{Alexei Nikolaenko}
\author{Denis Funfschilling}
\author{Guenter Ahlers}
\affiliation{Department of Physics and iQUEST,\\ University of
California, Santa Barbara, CA  93106}
\date{ \today} 

\begin{abstract}
We describe three apparatus, known as the large, medium, and small apparatus,  used for high-precision measurements of the Nusselt number $\cal N$ as a function of the Rayleigh number $R$ for cylindrical samples of fluid and present results illustrating the influence of the finite conductivity of the top and bottom plates on the heat transport in the fluid. We used water samples at a mean temperature of 40$^\circ$C (Prandtl number $\sigma = 4.4$). The samples in the large apparatus had a diameter $D$ of 49.69  cm and heights $L \simeq 116.33, 74.42, 50.61,$ and 16.52 cm. For the medium apparatus we had  $D = 24.81$ cm, and $L = 90.20$  and 24.76 cm. The small apparatus contained a sample with  $D = 9.21$ cm, and $L = 9.52$ cm. For each aspect ratio  $\Gamma \equiv D/L$ the data covered a range of a little over a decade of $R$. The maximum $R \simeq 1\times10^{12}$ with Nusselt numbers ${\cal N} \simeq 600$ was reached for $\Gamma = 0.43$. Measurements were made  with both Aluminum (conductivity $\lambda_p = 161$ W/m K) and Copper ($\lambda_p = 391$ W/m K) top and bottom plates of nominally identical size and shape. For the large and medium apparatus the results with Aluminum plates fall below those obtained with Copper plates, thus  confirming qualitatively the prediction by Verzicco that plates of finite conductivity diminish the heat transport in the fluid. The Nusselt number $\cal N_{\infty}$ for plates with infinite conductivity was estimated by fitting simultaneously Aluminum- and Copper-plate data sets to an effective powerlaw for ${\cal N}_{\infty}$ multiplied by a correction factor $f(X) = 1 - exp[-(aX)^b]$ that depends on the ratio $X$ of the thermal resistance of the fluid to that of the plates as suggested by Verzicco. Within their uncertainties the parameters  $a$ and $b$ were independent of $\Gamma$ for the large  apparatus and showed a small $\Gamma$-dependence for the medium apparatus. The correction was larger for the large, smaller for the medium, and negligible for the small apparatus.
\end{abstract}

\maketitle

\section{Introduction}
\label{sec:introduction}

The global heat transport by  turbulent Rayleigh-B\'enard convection (RBC) in a fluid heated from below  usually  is expressed in terms of the Nusselt number 
\begin{equation}
{\cal N} = Q L / \lambda \Delta T
\label{eq:N0}
\end{equation}
 where $Q$ is the heat-current density, $L$ the sample  height, $\Delta T$ the applied temperature difference, and $\lambda$ the thermal conductivity of the fluid in the absence of convection. The dependence of ${\cal N}$ on the Rayleigh number
\begin{equation}
R = \alpha g \Delta T L^3/\kappa \nu
\label{eq:R}
\end{equation}
($\alpha$ is the isobaric thermal expension coefficient, $\kappa$ the thermal diffusivity,  and $\nu$ the kinematic viscosity) and the Prandtl number $\sigma = \nu/\kappa$ is a central prediction of various theoretical models. \cite{Si94,Ka01,AGL02,GL00,GL01,GL02} 
 
One of the experimental problems in the measurement of ${\cal N}(R)$ is that the side wall often carries a significant part of the heat current.  \cite{Ah01,RCCHS01,Ve02,NS03} It is difficult to correct for this effect because the temperature field in the side wall is two-dimensional. Thus part of the applied current will pass from the wall near its bottom into the fluid and from the fluid near its top into the wall, thereby affecting the fluid flow. This effect can be minimized by using a fluid of relatively large conductivity confined by side walls of relatively low conductivity. Water confined by Plexiglas or Lexan is a good choice from this point of view. 
However, a different experimental problem associated with fluids of relatively large conductivity was pointed out recently by Chaumat et al, and by Verzicco. \cite{CCC02,Ve04} Using direct numerical simulation, Verzicco showed that the top and bottom end plates of the convection sample,  when they have a finite conductivity, diminish the heat transport in the fluid. This effect increases as the Nusselt number grows with increasing $R$ and, unless understood quantitatively,  severely limits the largest $R$ at which quantitative estimates of the idealized Nusselt number ${\cal N}_\infty(R)$ in the presence of plates with infinite conductivity can be obtained.  Verzicco found that, within his numerical resolution, the ratio $f(X)$ of the measured Nusselt number ${\cal N}$ to ${\cal N}_\infty$ depends only on the ratio $X$ (see Eqs.~\ref{eq:X_1} to \ref{eq:X_3} below) of the effective thermal resistance of the fluid to the thermal resistance of the end plates. He  derived an empirical form for $f(X)$ from his simulation. However, the numerical work was done using certain idealizations that are not fully realized in a typical experiment, and it can not be expected that Verzicco's result for $f(X)$, when used with experimental data for $\cal N$, will yield quantitatively reliable results for ${\cal N}_\infty$.

We used one old and built two new apparatus suitable for high-precision heat-transport measurements. The old one, to be called the small apparatus, was described before.\cite{small_app} It contained samples of aspect ratio near one with a diameter near 9 cm. The other two, called the medium and large apparatus, will be described in detail in the next section.  They were similar to each other in design, except that the large (medium) one  accommodated samples of diameter $D = 49.7$ cm ($D = 24.8$ cm). We describe the large apparatus in detail in this paper. Except for the length of the side wall, the medium one used identical vertical dimensions, but had radial dimensions reduced appropriately. 

Since the vertical structures of the end plates were nearly identical for the medium and large apparatus, we expected the same correction factor $f(X)$ to apply for the medium and large apparatus. However, the measurements revealed that this is not the case. We made measurements for several aspect ratios $\Gamma \equiv D/L$ with two types of top and bottom plates of nominally identical shape and size. One set of plates was made of copper with a conductivity $\lambda_{Cu} = 391$ W/m K and the other of aluminum with $\lambda_{Al} = 161$ W/m K. The results indeed showed a lower effective Nusselt number  for the Aluminum plates, and for a given $\Gamma$ the difference increased with increasing Rayleigh number. As expected, our data were not quantitatively consistent with the function $f(X)$ obtained by Verzicco.  Thus we used the Copper- and Aluminum-plate data at a given $\Gamma$ to derive an experimental correction function that, when applied to both data sets, caused all points to fall onto a single  curve. In this analysis we retained Verzicco's conclusion that the correction can be expressed as a function of the ratio $X$ of the thermal resistance of the fluid to that of the end plates. For the large apparatus we found that the correction function is, within our resolution,  independent of $\Gamma$; but for the medium apparatus the data yielded a small $\Gamma$-dependence. Contrary to our expectations $f(X)$ was closer to unity at a given $X$ for the medium apparatus than it was for the large one. This trend continued for the small apparatus, which yielded the same results within our resolution with copper and aluminum plates. For the medium and large apparatus we believe that the corrected results give the true Nusselt number, unencumbered by side-wall or end-plate effects, within a percent or so for Rayleigh numbers up to $R \simeq 1\times 10^{12}$. For the small apparatus a sidewall correction of less than 2 percent was required. 

Some of our results for ${\cal N}_\infty$ already have been published elsewhere.\cite{NA03,NBFA05}

\section{Apparatus and Performance}
\label{sec:apparatus}

\subsection{Hardware}
\label{sec:hardward}

\subsubsection{Large and Medium Apparatus}

Here we describe in detail the large apparatus. For the medium one horizontal dimensions are reduced by approximately 24 cm, and vertical dimensions are the same. A schematic diagram is shown in Fig.\ref{fig:apparatus}. From bottom to top, we find first a catchpan (A) capable of containing up to 200 liters of fluid in case of a leak. Supported above it is a 81 cm diameter support plate (B) made of high-strength aluminum alloy. It stands on three legs [solid lines between (A) and (B)] that consist of threaded 1.27 cm diameter steel rods screwed into tapped holes in part (B). The entire apparatus could be leveled by adjusting these legs. Part (C) was a bottom adiabatic shield made of aluminum. It was supported above part (B) by a 50 cm diameter steel cylinder  [vertical lines between (B) and (C)] with a height of 7.6 cm and a wall thickness of 0.32 cm.  The central area of 49.5 cm diameter of the bottom of (C) was covered uniformly by parallel straight  grooves of 0.76 cm depth and 0.40 cm width, interconnected by semicircles at their ends. Adjacent grooves were separated by 1.9 cm.
Epoxied into the grooves was a heater made of AWG No. 15 Nichrome C resistance wire surrounded by fiberglas sleeving. This heater had a total length of 10 m and a resistance of 6.8 $\Omega$.\cite{med_htr} A second auxiliary heater with a 7 $\Omega$ resistance was wound around the outside of the shield. 

Suspended above the shield, on a steel cylinder of 3.8 cm height and 0.32 cm wall thickness, was the bottom plate (D) of the sample. It was made in one case of high-strength aluminum alloy, and in the other of oxygen-free high-conductivity copper.  It had a thickness of 3.5 cm. Its top surface was finely machined, with tool marks of depth less than 3 $\mu$m. When aluminum was used, the plate was coated by the ``Tufram" process \cite{FNTuff}; the copper plate was uncoated. The bottom surface contained the same type of Nichrome heater as the bottom shield. Its diameter was 54.6 cm. A top central section had a reduced diameter of 49.5 cm which was a close slide fit into the plexiglas side wall cylinder (E). At one point of the side of the section of reduced diameter there was a vertical 0.16 cm diameter semi-circular groove through which the fluid could enter the system. Five small holes were drilled from below at an angle into the bottom plate to within 0.32 cm of its top surface, and thermistors were mounted in these holes. 

The inner diameters $D$ of the plexiglas side walls (E) were measured at several angular and axial positions, in several cases both before and after use. The standard deviation from the mean of the measurements for a particular side wall generally was 0.03 cm or less. For the large apparatus, where several side walls were used, the mean values of $D$ were remarkably constant from one aspect ratio to another, and all fell in the range  $49.68 < D < 49.70$ cm.  For the medium apparatus we had $D = 24.81$ cm.  We saw no statistically significant evidence of changes due to the exposure to water during the experimental runs.  We estimate that systematic errors from uncertainties in the cross sectional area were less than 0.1\%. The wall thickness was 0.63 cm for the large and 0.32 cm for the medium apparatus. The wall lengths $L$ were uniform around the circumference to $\pm 0.01$ cm and determined the aspect ratio of the sample. Thus the uncertainty in the geometry due to uncertainties in $L$ also was less than 0.1\%. In the present paper we report on measurements for $L = 116.33, 74.42, 50.61$, and 16.52 cm, corresponding to $\Gamma = 0.427, 0.667, 0.981,$  and 3.008, for the large apparatus. For the medium one we used $L = 90.20$ cm and 24.76 cm, corresponding to $\Gamma = 0.275$ and 1.00.  
We deliberately did not use  an aspect ratio close to $1/2$ in an effort to avoid the multi-stability reported for this value on  the basis of experiment \cite{RCCH04} as well as of numerical simulations \cite{VC03}.

The side wall  extended 2.38 cm below the top surface of the bottom plate. An ethylene-propylene O-ring sealed the system from the outside well below the top surface of the bottom plate. A similar construction was used to terminate the side wall at the top. This construction minimizes the heat flow into the wall and provides a well defined geometry for simulations of this heat-flow problem. \cite{Ah01} 

The side wall was surrounded by an adiabatic side shield (F) made of aluminum. Epoxied to the outside of this shield was a double spiral consisting of 15 m of aluminum tubing. Water from a temperature-controlled circulator flowed through the tubing. The shield was suspended above the support plate (B) by six 1.9 cm diameter and 17.5 cm long plastic rods (not shown in Fig.~\ref{fig:apparatus}). During measurements its temperature was kept close to the mean temperature of the system.

The top of the sample was provided by an aluminum or copper top plate (H) which was similar to the bottom plate in its dimensions. The aluminum plate was also ``Tufram" plated. The top plate contained an outlet for the fluid which was identical to the inlet in the bottom plate, but in the sample assembly care was taken to locate the outlet at an angular position opposite to the inlet. The top plate differed from the bottom one in that it did not contain a heater. Instead, it was cooled by temperature controlled water from a refrigerated circulator.\cite{FNNeslab} It had a thickness of 3.34 cm, and a double-spiral water-cooling channel was machined directly into it. The channel width and depth were $0.95\times 2.54$ cm$^2$. The spacing between adjacent turns of the spiral was 2.54 cm.\cite{med_spiral} The bottom of the groove came within 0.79 cm of the metal-fluid interface. An additional plate was ``O"-ring sealed to the top plate from above to close the spiral channel. Small holes were drilled through the two-plate composite from above to within 0.32 cm of the aluminum-fluid interface, and calibrated thermistors were mounted with their heads within 0.48 cm of the convecting fluid. Each thermistor was protected from circulating water by an additional small ``O"-ring between the two plates.
To avoid convection of air in the vicinity of the sample,  the entire space outside the sample but inside the dashed rectangle K was filled with low-density (firmness rating 1) polyurethane foam sheet.

The large apparatus was intended also for future work with compressed gases as the fluid. Thus, a retaining plate (I) and tension rods (G) were provided to sustain the force exerted by pressures up to 10 bar. They were not in use during the measurements reported here.       

The medium and large apparatus each  contained 16 thermistors \cite{Fenwal} that were calibrated simultaneously in a separate apparatus against  a laboratory standard based on a standard platinum thermometer. Deviations of the data from the fit were generally less than 0.002 $^\circ$C.  
In the top as well as the bottom plate one thermistor terminated at the plate center  and four were located equally spaced on a circle of 43 cm diameter. The remaining thermistors were mounted on the adiabatic bottom shield, the adiabatic side shield, and the outside of the plexiglas side wall. 
The power of the bottom-plate heater was determined by a four-lead method.

The sample was leveled to better than 0.1$^\circ$. In order to look for any influence of a misalignment relative to gravity, a few data points were taken also with the sample tilted by 2$^\circ$.  No significant effect on $\cal N$ was noted.

\subsubsection{Small Apparatus}

The small apparatus had been used extensively for the study of pattern formation near the onset of RBC, and was described in detail in Ref.~\cite{small_app}. The top plate was cooled from above by circulating temperature-controlled water. The bottom plate was heated by a metal-film heater from below. The apparatus was used for the work reported in Refs.~\cite{XBA00,Ah01,AX01}, and there it  contained a sample of acetone at a mean temperature of 32.00$^\circ$C ($\sigma = 3.97)$. A recent new set of measurements was made as well with this configuration. The sample had a 0.318 cm thick sapphire top plate and a 0.635 cm thick Aluminum bottom plate. \cite{XBA00,Ah01,AX01} The transparent sapphire had the advantage of giving optical access. It had the disadvantage that thermometers could not be mounted within it and that the top temperature had to be determined with a thermometer immersed in the  cooling bath. A small correction for the series resistance of the sapphire and of a boundary layer above the sapphire in the water bath was then required. \cite{FNBL} High-density polyethylene side walls compatible with acetone, were used. They had a relatively high conductivity and the fluid had a relatively small conductivity. Thus a significant wall correction, ranging from about 15\% at small to about 8\% at large $R$,  was required \cite{Ah01} and introduced an uncertainty of several percent for the final corrected data.  In recent measurements the sapphire top and aluminum bottom plates were replaced with 0.635 cm thick Copper plates with thermometers mounted directly in them, within 0.3 cm of the metal-fluid interface. Corrections for the temperature differences between the thermometers and the metal-fluid interfaces were then quite small. As shown in the Appendix, the results demonstrate that no Verzicco correction is required for the runs with the sapphire top and aluminum bottom plates. 

New measurements were made also in a different configuration, using water at 40$^\circ$C as the fluid. Here top and bottom aluminum or copper plates of identical dimensions were used. The plates were 1.90 cm thick. Of this thickness, 0.63 cm protruded as an anvil into a 9.21 cm diameter Lexan side wall. The side wall was ``O'-ring sealed to the plates at a point beyond the metal-fluid interface, well away from the bulk fluid.The side wall had a thickness of 0.32 cm. Only a  small wall correction was required.  First, a correction was made for the conductance of the system with an evacuated sample cell. This correction was dominated by a contribution due to the conductance from the bottom plate directly to a surrounding water bath.\cite{small_app} Based on model 2 of Ref.~\cite{Ah01} we estimate that the actual non-linear correction for the wall conduction was about 1.7\% at $R \simeq 3\times10^7$ where ${\cal N} \simeq 24$ to 0.8\% for $R \simeq 10^9$ where ${\cal N} \simeq 66$. We found that the results with copper and aluminum plates agreed with each other within their uncertainty, indicating that the Verzicco correction was not required in the Rayleigh-number range and for the geometry of this apparatus. We believe that the wall-corrected results are subject to systematic errors of less than one percent.

\subsection{Procedure and Performance}
\label{sec:performance}

This section is based on experience with the large apparatus. The performance of the medium one was similar. 

The measurements were made using de-ionized water. The water was degassed by circulating it with a peristaltic pump for one or two days through a small stirred container external to the sample at about 50$^\circ$C. After this procedure, visual inspection of the sample showed complete absence of any bubbles. All data points are for a mean temperature close to 40$^\circ$C where the Prandtl number $\sigma$ is 4.38. The variation of $\sigma$ over the applied temperature difference can be estimated from $(1/\sigma)(d\sigma/dT) = 0.020$K$^{-1}$. At 40$^\circ$C water has the following properties: density $\rho = 992.2$ kg/m$^3$, isobaric thermal  expansion coefficient $\alpha = 3.88\times 10^{-4}$ K$^{-1}$, thermal conductivity $\lambda = 0.630$ W/m K, heat capacity per unit mass at constant pressure $C_P = 4170$ J/kg K, thermal diffusivity $\kappa = 1.528\times 10^{-7}$ m$^2$/s, and kinematic viscosity $\nu = 6.69\times 10^{-7}$ m$^2$/s.

The system was equilibrated with both top and bottom plates at 40$^\circ$C for over a day. Thereafter the power needed to maintain the bottom-plate temperature was constant and equal to 0.2 W. We do not know the origin of this parasitic heat loss,  but subtracted it from all subsequent measurements. It was about 0.8\%  (0.04\%) of $Q$ for $\Delta T = 1 (10) ^\circ$C. A possible dependence of this correction on the temperature of the bottom plate constitutes one of the possible sources of error for  the measurements with small $\Delta T$.

We determined the conductance between the bottom plate and the bottom adiabatic shield by changing the shield temperature with the bottom plate at 40$^\circ$C, and found it to be about one W/$^\circ$C. Thus, the regulation of the shield temperature  within better than 0.01$^\circ$C of the bottom-plate temperature reduced heat exchange between the shield and the bottom plate to a negligible value.

We measured the influence of the adiabatic side-shield temperature $T_s$ on the heat current necessary to maintain a constant temperature difference across the sample. We found this current to decrease by 1.3 W when $T_s$ was increased by 1$^\circ$C. The side shield was always kept within about 0.1$^\circ$C of the mean sample temperature, and thus heat exchange with it was not a significant source of error except perhaps at the smallest values of $\Delta T$ where the heat current through the sample was small.

In the remainder of this section we describe the performance of the system for $\Gamma \simeq 1$. Similar behavior was found for other $\Gamma$ values.

Data points were obtained by holding constant both the top and bottom plate temperature. For the bottom plate this was done by adjusting the heater power in a digital feedback loop using the central thermometer ($T_0$). For the top plate a constant temperature was achieved by circulating temperature-controlled water \cite{FNNeslab}  through the double spiral. The heat current applied to the bottom plate and all five  thermometers in each plate were monitored for a minimum of 8 hours, frequently for one day, and on occasion for ten days under nominally constant external conditions. Usually a constant current and constant temperatures were obtained after less than an hour. This is illustrated in Fig.~\ref{fig:Neff}, where we show the apparent Nusselt number computed from the top- and bottom-plate temperatures and the heat current during the transient after a change of the Rayleigh number from about $1\times10^{10}$ to $8.4\times 10^{10}$. One sees that the  transients had a relaxation time $\tau_N$ of about 300 sec, and died out within our resolution after only about half an hour. The relaxation time was about the same for the Rayleigh number (i.e. for $\Delta T$), and we believe it to be controlled by the rate at which we can provide/remove the energy necessary to heat/cool  the bottom/top plate rather than by hydrodynamic transients. Nonetheless it is instructive to compare the observed relaxation time with an estimate of an internal time constant of the system. It is significantly faster than a simple relaxation process involving the entire system and an effective vertical thermal diffusion time $\tau_N = L^2/(\pi^2 {\cal N}\kappa) \simeq 600 s$. This suggests that the  relatively thin boundary layers adjacent to the top and bottom plates equilibrate quite quickly, and that the large-scale circulation is also established on a relatively short time scale. Our experience indicates that the concerns expressed by Roche et al. \cite{RCCH04} about very long time constants for Nusselt-number measurements using water are unwarranted, at least for the aspect ratios that we examined. 

The currents and temperatures recorded after the first several hours were time averaged. The five averaged temperatures for each plate were averaged to give our best estimate of the top and bottom plate temperatures. A correction to these temperatures was made for the temperature change across the aluminum or copper layers  between the fluid and the thermistor heads. For the aluminum plates this correction was estimated to be 0.8\%  (1.5\%) of $\Delta T$ for $\Delta T = 1 (10) ^\circ$C; for the copper plates it was smaller by a factor of 2.4. The results for $T_t$ and $T_b$ yielded the final values of $\bar T = (T_b + T_t)/2$, and of the value of $\Delta T = T_b - T_t$ use to compute $\cal N$.

Recently it was appreciated that the heat flux from (to) the bottom (top) plate occurs preferentially near the circumference of the plate, with smaller fluxes near the center. \cite{SQTX03} This will inevitably lead to horizontal thermal gradients in plates of finite conductivity. To our knowledge no systematic measurements of such gradients have been made for actual experimental conditions. In Figs.~\ref{fig:dT/DT_Al} and \ref{fig:dT/DT_Cu} we show the horizontal temperature differences $T_i - T_0$, $i = 1, 2,3,4$ between thermometers 1 to 4 near the plate periphery and thermometer 0 at the plate center, normalized by the applied vertical temperature difference $\Delta T$. Here Fig.~\ref{fig:dT/DT_Al} is for the Aluminum and Fig.~\ref{fig:dT/DT_Cu} is for the Copper plates. In both cases $(T_i - T_0)/\Delta T$ is at most of the order of a few percent. Surprisingly, the results for Copper (which has 2.5 times the conductivity of Aluminum) are not very different in magnitude from the Aluminum data. We note that in detail the results will depend on the nature and orientation of the large-scale circulation which prevails in the sample. For the determination of the Nusselt numbers we used the average of all the thermometers in a given plate. We would estimate that this horizontal inhomogeneity of the temperature at the fluid boundary could introduce a systematic error of order 1 percent into the results for $\cal N$.

\section{Effect of the Top- and Bottom-Plate Conductivities}
\label{sec:results}

\subsection{Experimental evidence}
\label{sec:uncorr}
  
In Fig.~\ref{fig:loglog} we show the $\Gamma = 1$ results (large apparatus) for aluminum (copper) plates as open squares (circles) on double logarithmic scales. Corresponding data were obtained in the large apparatus for $\Gamma = 3.00$, 0.67, and 0.43, in the medium apparatus for $\Gamma = 0.275$ and 1.00, and in the small apparatus for $\Gamma = 0.967$. Also shown, as open triangles, are the data for acetone (with Prandtl number $\sigma = 4.0$) from Ref.~\cite{XBA00}. The acetone data were corrected for the wall conduction using model 2 of Ref.~\cite{Ah01} (see also Ref.~\cite{AX01}). Their range of $R$ only barely overlaps with that of the present data, but there is consistency. The solid line is the prediction of the model by Grossmann and Lohse (GL). This model had been fitted to data that included the acetone results, and thus it fits the acetone data quite well. The new data with aluminum plates depart from the prediction more and more as $R$ increases. The results with copper plates agree well with the GL prediction. However, as we shall see below, to some extent this agreement is illusory because the agreement will be spoiled somewhat by a correction for the finite conductivity of the copper plates. In Sect.~\ref{sec:Verzicco} we show that  the difference between the experimental results obtained with copper and aluminum plates can be understood in terms of the finite plate-conductivity as proposed by Verzicco. \cite{Ve04}

\subsection{Verzicco Correction}
\label{sec:Verzicco}

Recently it was shown by Verzicco \cite{Ve04}, by including the top and bottom plates in direct numerical integrations of the Boussinesq equations, that a finite conductivity of the top and/or bottom plates of a sample will diminish the heat transport by the fluid. The physical reason for this phenomenon is found in the significance of ``plumes" for the heat transport. Much of the thermal energy that leaves the more or less quiescent boundary layer above the bottom plate does so in the form of modestly sized relatively warm and thus buoyant fluid volumes that have become known as plumes. These plumes leave behind an enthalpy deficiency that takes the form of a ``cold spot" in the bottom plate. This cold spot, if it has a finite lifetime, decreases the likelihood of further plume formation, and thus decreases the heat flux out of the bottom boundary layer. An analogous process occurs at the cold top plate. 

Verzicco carried out extensive numerical simulations for a system of aspect ratio $\Gamma = 1/2$, with identical top and bottom plates and isothermal boundary conditions on the dry sides of the plates. For that case he showed that the effect of the plate conductivity on the measured Nusselt number can be described in terms of a function $f(X)$, with
\begin{equation}
{\cal N} = f(X) {\cal N}_\infty\ .
\label{eq:n_infty}
\end{equation}
where ${\cal N}_\infty$ is the ideal Nusselt number in the presence of perfectly conducting rigid top and bottom plates. The argument $X$ is the ratio of the thermal resistance of the fluid to that of an end plate, i. e. 
\begin{eqnarray}
X &=& R_f/R_p \label{eq:X_1}\\
&=& X_0/{\cal N}\ ; \label{eq:X_2}\\
X_0 &\equiv & \lambda_p L / (\lambda_f e) \label{eq:X_3}
\end{eqnarray}
with $R_f$ and $R_p$ equal to the resistances of the fluid and of a plate respectively. Here $\lambda_p$ and $\lambda_f$ are the end-plate and fluid conductivities respectively,  and $e$ is the thickness of one plate. 
Several of the conditions used in the simulations do not correspond to the experiment. The actual bottom plate is closer to experiencing a constant heat  flux, the top plate is cooled in its interior by water channels, and the plate geometries are significantly different at the top and bottom. Further, the experiments are not for the same $\Gamma$ as the simulations. Thus we do not expect the Verzicco prediction to be quantitatively applicable; but the difference between the results with copper and aluminum plates shown in Fig.~\ref{fig:loglog} is in the predicted direction and of about the right size. Thus the measurements provide qualitative confirmation of the importance of the Verzicco effect.

At large $R$ the dependence of $\cal N$ on $R$ can be written approximately as ${\cal N} = N_0 R^{1/3}$. We note that in this approximation $X$ is given by

\begin{equation}
X = \left[\frac {\lambda_p (\kappa \nu)^{1/3}}{\lambda_f e N_0 (\alpha g)^{1/3}}\right ]\Delta T^{-1/3}
\label{eq:XofDT}
\end{equation}

\noindent and is independent of the height and diameter of the sample. Thus, regardless of the aspect ratio and the overall dimensions, the correction is predicted to be nearly the same for the same $\Delta T$. Approximately, it should depend only on a constant combination of sample cell and fluid properties, and on $\Delta T^{-1/3}$. As we shall discuss below, we find this not to be the case. The correction for the large apparatus is significantly larger than it is for the medium one even though the top and bottom plates have the same vertical dimensions. For the small apparatus (which did have a different end-plate structure) no correction was needed.

From his numerical results Verzicco derived the emperical formula
\begin{equation}
f(X) = f_V(X) = \{1 - exp[-(X/4)^{1/3}]\} X / (X - 2)
\label{eq:f_V}
\end{equation}
for the case where the Nusselt number is determined from the temperatures at the fluid-plate interfaces (as is the  case in the experiment). In Fig.~\ref{fig:f_of_X} we show this function as a dashed line. 

\subsection{Results for the large apparatus}

To obtain an estimate of the size of the correction, we evaluated $X$ as a function of $R$ for four aspect ratios using an average $e \simeq 1.9$ cm of the thickness of the bottom plate and of the part of the top plate below the water cooling channels. Using this and the measured Nusselt numbers, we obtained the values shown in Fig.~\ref{fig:X} as solid and dashed lines for copper and aluminum respectively for the large apparatus. Clearly the copper plates, having much larger values of $X$,  should yield results much closer to ${\cal N}_\infty$ than the aluminum plates.

In Fig.~\ref{fig:comp1}a we present experimental results for ${\cal N}$. We use the ``compensated" form ${\cal N}/R^{0.3}$ in order to obtain sufficient resolution in the graph. We compare the uncorrected results for $\cal N$ (open symbols) for $\Gamma = 3.00$ (left data set), $\Gamma = 1.00$ (middle data set)  and $\Gamma = 0.67$ (right data set) with the estimates of ${\cal N}_\infty$ based on Eq.~\ref{eq:f_V} and the values of $X$ given in Fig.~\ref{fig:X} (solid symbols). One sees that the functional form Eq.~\ref{eq:f_V} does not lead to a collapse of the data obtained with the copper and aluminum plates, indicating as expected that the Verzicco function $f_V$ is not quantitatively applicable to the data. 

In order to obtain a better estimate of ${\cal N}_\infty$, we replaced $f_V(X)$ with the slightly different emperical function
\begin{equation}
f(X) = 1 - exp[-(a X)^{b}]\ .
\label{eq:f}
\end{equation}
On the basis of the physical process envisioned by Verzicco (see Sect.~\ref{sec:Verzicco}) we feel that the form of Eq.~\ref{eq:f} is a better choice for two reasons. First it is a monotonically increasing function of $X$, which it should be since we expect the correction to become smaller as $X$ becomes larger ($\cal N$ becomes smaller, see Eq. 5). This is so because we expect fewer plumes to be emitted for smaller $\cal N$. Second, $f(X)$ is less than  unity for all values of its argument which it should be because the physical process envisioned can not {\it enhance} the heat transport by the fluid. We note that $f_V$ has a minimum at small $X$ and is greater  than unity both for small and large $X$ (of course we understand that it was meant to be used only over a limited range of $X$ where it was monotonically increasing and less than unity).  

We simultaneously fitted the aluminum- and copper-plate data for a given $\Gamma$ to the equation
\begin{equation}
{\cal N} = N_0 R^{\gamma_{eff}} f(X)\ ,
\label{eq:N}
\end{equation}
i.e. over the experimental range of $R$ for a given $\Gamma$ (about one decade) we represented ${\cal N}_\infty$ by an effective powerlaw with an effective exponent $\gamma_{eff}$. The parameters $N_0$, $\gamma_{eff}$, $a$, and $b$ were least-squares adjusted. The results are listed in Table~\ref{table:f_fit}. Typical statistical errors for both $a$ and $b$ were near 0.02, but systematic errors due to small uncertainties in the sample diameter and height made a similar contribution to their uncertainties. Thus, within experimental error we found that $f(X)$ is independent of $\Gamma$. We adopted the values $a = 0.275$ and $b = 0.39$ for all $\Gamma$. The corresponding $f(X)$ is shown as a solid line in Fig.~\ref{fig:f_of_X}.  We repeated the least-squares fits, adjusting $N_0$ and $\gamma_{eff}$. The resulting parameter values are again given in Table~\ref{table:f_fit}. Corresponding results for ${\cal N}$ and ${\cal N}_\infty$ based on the copper- and aluminum-plate data are shown in Fig.~\ref{fig:comp1}b. One sees that the results for ${\cal N}_{\infty}$ (solid symbols) for aluminum (squares) and copper (circles) plates within their scatter collapse onto a single curve. 

\subsection{Results for the medium apparatus}

Estimates of $X$ for the medium apparatus are shown  in Fig~\ref{fig:X} as dash-dotted and dotted lines for copper and aluminum plates respectively. They are seen to be comparable in size to those for the large apparatus. Results of measured values of ${\cal N}/R^{0.3}$ for $\Gamma = 1.00$ and 0.275 are shown in Fig.~\ref{fig:comp2} as open symbols, with the squares (circles) obtained with aluminum (copper) plates. Again there is a difference that can be attributed to the difference in the plate conductivity; but this difference is smaller than it was in the case of the large apparatus. Again a fit of Eq.~\ref{eq:f} to the data yielded values of ${\cal N}_\infty$ (solid symbols) that agreed for the two types of plates. The coefficients obtained from the fits are given in Table~\ref{table:f_fit}. They differ slightly, but significantly, for the two aspect ratios. The corresponding functions $f(X)$ are shown in Fig.~\ref{fig:f_of_X} as dash-dotted lines. These lines lie well above the solid one, reflecting the fact that the correction for the medium apparatus is much smaller than it is for the large  apparatus. 

\subsection{Results for the small apparatus}

Results of measured values of ${\cal N}/R^{0.3}$ for $\Gamma = 0.967$ obtained with the small apparatus using water at 40$^\circ$C as the fluid are shown in Fig.~\ref{fig:comp3} as open symbols, with the squares (circles) obtained with aluminum (copper) plates. In this case we see that there is no significant influence of the plate conductivity on the results. Thus either set of data should be a good representation of ${\cal N}_\infty$.

\section{Summary and Conclusion}

In this paper we described a large, a medium, and a small apparatus for the measurement of heat transport in turbulent Rayleigh-B\'enard convection. Over the range $10^8 \leq R \leq 10^{12}$ we studied the influence of the finite conductivity of the top and bottom confining plates by making measurements in each apparatus for samples with various aspect ratios. In each case we used  two sets of plates of nominally identical geometry, one made of copper with relatively high conductivity and the other made of aluminum with lower conductivity. As predicted by Verzicco, \cite{Ve04} we found for the large and medium apparatus that the Nusselt number was diminished more when the plate conductivity was smaller, and that this effect could be represented by an empirical correction factor $f(X)$ that is a function of the ratio $X$ of the thermal resistance of the plates to that of the fluid (Eq.~\ref{eq:n_infty}). For the large apparatus we found that $f(X)$ is, within our resolution, independent of the aspect ratio of the sample (see Eq.~\ref{eq:f} and Table~\ref{table:f_fit}); for the medium apparatus there was a small aspect-ratio dependence. Although both the medium and large apparatus required a plate correction to obtain the conductivity ${\cal N}_\infty$ for plates of infinite conductivity, we found that this correction was smaller for the medium apparatus. For the small apparatus the data indicate that no correction is required. 

One aspect of this phenomenon is worth noting. In the $R$-range of interest ${\cal N}$ is approximately proportional to $R^{1/3}$. In this approximation one can show  (see Eq.~\ref{eq:XofDT}) that, at constant plate thickness $e$,  the argument $X$ of $f(X)$ is proportional to $\Delta T^{-1/3}$ and independent of the sample size and aspect ratio. Thus, to a good approximation the correction for the finite plate-conductivity is expected to be equally important for measurements at relatively small $\cal N$ and $R$, provided that the sample height is relatively small so that the applied temperature difference $\Delta T$ is large. This is illustrated above by the data in Fig.~\ref{fig:comp1} for various $\Gamma$ (and thus $L$) but with the same diameter $D$ and plate thickness $e$. 

From the measurements for the large and medium sample we found that a change of the overall size of the sample (i.e. of $L$ and $D$ in proportion to each other) at constant $\Gamma$ and $e$ changes the correction $f(X)$ considerably even though the argument $X$ is about the same. 
The dependence of $f(X)$ on the physical size of the apparatus seems surprising. At this point we can only speculate about a possible reason. One explanation might be found by considering again the limit where (at constant $\Delta T$) ${\cal N} \sim R^{1/3} \sim L$ (see Eq.~\ref{eq:R}) and where (see the previous paragraph)  $X$ is independent of $L$. In that limit we consider the size ${\it l}_P$ of a ``plume" that created a thermal ``hole" of lateral extent ${\it l}_P$ in, say, the bottom plate. We expect ${\it l}_P$ to be of the same order as the thermal boundary-layer thickness ${\it l} \simeq L/2{\cal N}$ \cite{Si94}, which (at constant $\Delta T$) yields ${\it l}_P$ independent of $L$. Finally, we consider the time $\tau_P = {\it l}_P/v$ needed by the large-scale circulation (LSC) of speed $v$ to traverse the thermal hole. If this time is long compared to the thermal relaxation time of the hole in the bottom plate (which should depend only on the plate properties), then the effect of a previously generated plume on further plume emission should be small. For $v$ we have $v \propto R_e \nu/L$ where $R_e$ is the Reynolds number of the LSC. To a good approximation one expects \cite{Si94,GL00} $R_e \sim R^{4/9} \sim L^{4/3}$, and thus   $\tau_P \sim L^{-1/3}$. We conclude that indeed $\tau_p$ is larger for a shorter cell, thus suggesting a diminished Verzicco effect. This argument could be affected by an $R$ dependence of the plume density above (below) the bottom (top) plate; little is known about this from experiment, but a recently proposed model \cite{GL04} suggests that this density is independent of $R$.

Another unexpected result is the noticeable, albeit small,  $\Gamma$-dependence of $f(X)$ for the medium sample. Here we note that the LSC structure for the two aspect ratios most likely is very different. For $\Gamma \simeq 1$ we expect a single role filling the entire cell. For $\Gamma \simeq 0.28$ it is likely that two or more rolls positioned vertically above each other prevail. One would expect such a difference in the LSC geometry to influence the speed $v$ of the LSC above (below) the bottom (top) plates, leading to somewhat different sizes of the Verzicco effect. 

The work presented here enables us to make reliable estimates of the theoretically relevant Nusselt number ${\cal N}_\infty$ of a system with infinite thermal conductivity of the top and bottom boundaries. Some of the results for ${\cal N}_\infty$ have already been published elsewhere,\cite{NA03, NBFA05} and we shall discuss the implications of the results for  ${\cal N}_\infty$ in further detail in a future publication. Here we just note that ${\cal N}_\infty$ differs significantly from the predictions of the Grossmann-Lohse model \cite{GL01} with the model parameters currently in use. This can be seen in Fig.~\ref{fig:comp1} where the model prediction is shown as a solid line. It will have to be determined whether the model parameters can be adjusted so as to re-gain satisfactory agreement.

A second point to be noted is that the data in Figs.~\ref{fig:comp1} and  \ref{fig:comp2} allow at most a very small dependence of ${\cal N}_\infty$ on the aspect ratio $\Gamma$.  We appreciate that this observation differs from prior measurements \cite{XBA00,WL92} of $\cal N$ that were not corrected for the wall effect \cite{Ah01,RCCHS01,Ve02,NS03} or the Verzicco end-plate effect discussed in the present paper.

Finally, in an Appendix we re-examine previously published data and present new measurements with acetone as the fluid. We conclude that the older results had not been influenced significantly by the finite plate-conductivity effect.

\section{Acknowledgment}

This work was supported by the US Department of Energy through Grant  DE-FG02-03ER46080.

\section{Appendix}

Previous measurements, for a Prandtl number close to four, were reported in Refs.~\cite{XBA00}, \cite{Ah01}, and \cite{AX01} where acetone with $\sigma = 3.96$ was used. Those data covered the range $10^8 \leq R \leq 10^{10}$ and were corrected for the influence of the side wall on the heat flux, \cite{Ah01} but not for the effect of the end plates on the turbulent flow (which had not been anticipated at that time). Here we consider briefly  the end-plate effect. 

The sample cell had an inside diameter of 8.74 cm and a side wall of high-density poly-ethylene (HDPE) of thickness 0.38 cm. However, the top and bottom of the side wall each had a HDPE flange of thickness 0.73 cm (including the side-wall thickness) and heigh 0.47 cm adjacent to the fluid, thus making any estimate of the nonlinear side-wall correction \cite{Ah01} quite uncertain. The sample cell consisted, from bottom to top,  of an aluminum bottom plate of thickness $e_{Al} = 0.635$ cm (conductivity $\lambda_{Al} = 161$ W/m K), a sample of acetone of height $L = 8.70$ cm ($\lambda_{fl} = 0.159$ W/m K), a sapphire top plate of thickness $e_{sapp} = 0.317$ cm ($\lambda_{sapp} = 37$ W/m K), and a circulating water cooling  bath. The bottom-plate temperature was measured by thermistors imbedded in it about 0.2 cm from the fluid-solid interface; a correction for the temperature drop between the thermistors and the fluid  was negligible. The temperature at the top was measured in the water coling bath. The  temperature differences were corrected for the temperature drop across the sapphire (which ranged from 1 percent near $R=10^8$ to 3 percent near $R = 10^{10}$), and for the temperature drop due to an estimated thermal resistance $R_b \simeq 0.009$ K/W of a boundary layer in the water bath above the sapphire (which ranged from 0.4 percent near $R=10^8$ to 1.3 percent near $R = 10^{10}$). \cite{FNBL} A correction of 0.085 W/K (22 percent near $R = 10^8$ and 6.5 percent near $R = 10^{10}$) was measured with an evacuated sample cell and used to correct the heat current, but for present purposes no further (non-linear) correction for the side-wall effect \cite{Ah01} was made. The results for the reduced Nusselt number are shown in Fig.~\ref{fig:acetone} as open circles. 

To confirm the above results we made a new set of measurements after the sample had been taken apart and then assembled again four years later, and obtained results represented by the open squares in Fig.~\ref{fig:acetone}, in good agreement with the previous data. 

One might attempt to correct for the top- and bottom-plate influence using Eqs.~\ref{eq:n_infty} and Eq.~\ref{eq:f} with the coefficients given in Table~\ref{table:f_fit}. The nature and geometry of the boundaries in the acetone experiments were very different from those of the copper and aluminum plates used in the present work, and there is no reason why the results Eqs.~\ref{eq:n_infty} and Eq.~\ref{eq:f} with the coefficients in Table~\ref{table:f_fit} should apply quantitatively. Nonetheless it is instructive to examine the size of the predicted effect. We estimated an average plate resistivity $R_p = (e_{sapp}/\lambda_{sapp} + e_{Al}/\lambda_{Al})/2 = 0.62 $ cm$^2$K/W and computed $X$ from Eqs.~\ref{eq:X_1} to \ref{eq:X_3}. This model then predicts a modest correction to $\cal N$ (4 percent near $R = 10^{10}$) at the largest $R$.

To obtain an experimental estimate of the plate effect, we constructed a new sample cell.  The same HDPE side wall was used. The  top and bottom plates were replaced by copper plates of dimensions equal to those of the previous aluminum and sapphire plates.  In this case the top-plate temperature was measured with a thermistor imbedded in the top plate, thus eliminating any series resistance in the water bath. An estimate similar to that of the previous paragraph predicts that any plate correction should be completely negligible. The results for $\cal N$ are shown in reduced form in Fig.~\ref{fig:acetone} as solid circles. They agree well with the measurements with the aluminum bottom and sapphire top plate. Thus we conclude that the end-plate correction for the original acetone experiment was negligible. 

\vfill\eject

\begin{table}
\caption{Parameters obtained from fits of Eqs.~\ref{eq:f} and \ref{eq:N} to the data with aluminum and copper top and bottom plates. Entries marked with a $^*$ indicate fixed values.}
\vskip 0.1in
\begin{tabular}{cccccc}
$D (cm) $ & $\Gamma$ & a & b  & N$_0$ & $\gamma_{eff}$\\
\tableline
49.7 & 0.43  	& 0.240  &    0.434    & 0.06401  & 0.3303\\
49.7 & 0.67 & 0.250 & 0.428      & 0.0684 & 0.3281\\
49.7 & 1.00	& 0.288 & 0.361      & 0.0733 & 0.3261\\
49.7 & 3.00	& 0.279 & 0.392      & 0.0902 &0.3164\\
\tableline
49.7 & 0.43	& 0.275$^*$ & 0.39$^*$ & 0.06395  & 0.3306\\
49.7 & 0.67	& 0.275$^*$ & 0.39$^*$ & 0.0680 & 0.3287\\
49.7 & 1.00	& 0.275$^*$ & 0.39$^*$ & 0.0744 & 0.3251\\
49.7 & 3.00	& 0.275$^*$ & 0.39$^*$ & 0.0897 & 0.3167\\
\tableline
24.8 & 0.275 & 0.304 & 0.506 & 0.0838 & 0.3194\\
24.8 & 1.00   & 0.378 &  0.488 &  0.1047&  0.3101\\
\end{tabular}
\label{table:f_fit}
\end{table}

\begin{figure}
\caption{Schematic diagram (not to scale) of the large apparatus. The various parts are explained in the text.}\label{fig:apparatus}

\caption{The effective Nusselt number as a function of time  after a sudden change of the top and bottom plate temperatures.}\label{fig:Neff}

\caption{The horizontal temperature differences $(T_i - T_0)/\Delta T$ for the aluminum top (a) and bottom (b) plates and $\Gamma = 1$. Here $\Delta T$ is the vertical applied temperature difference, and $i = 1, 2,  3, 4$ labels the peripheral thermometers in counterclockwise sequence when viewed from above. Open circles: $i = 1$, solid circles: $i = 2$; open squares: $i = 3$; solid squares: $i = 4$.}\label{fig:dT/DT_Al}

\caption{The horizontal temperature differences $(T_i - T_0)/\Delta T$ for the copper top (a) and bottom (b) plates and $\Gamma = 1$. The symbols are as in Fig.~\ref{fig:dT/DT_Al}.}\label{fig:dT/DT_Cu}

\caption{The Nusselt number as a function of the Rayeigh number for $\Gamma = 1.0$ on logarithmic scales. Open triangles: Data for acetone from Ref.~\protect \cite{XBA00} after correction for the wall conduction. Open squares: Results from the large apparatus with aluminum top and bottom plates. Open circles: Results from the large apparatus with copper top and bottom plates. Solid line: the prediction of Ref.~\protect \cite{GL01}.}\label{fig:loglog}

\caption{The ratio $X$ of the thermal resistance of the fluid to the average resistance of the top and bottom plate as a function of the Rayleigh number $R$. Solid lines: large sample, copper plates. Dashed lines: large sample, aluminum plates.  From left to right, those lines are for $\Gamma = 3$, 1, 0.67, and 0.43. Dash-dotted lines: medium sample, copper plates. Dotted lines: medium sample, aluminum plates.  From left to right, those lines are for $\Gamma = 1$ and 0.275. The lines cover the range of $R$ used in the experiments.}\label{fig:X}

\end{figure}

\begin{figure}

\caption{Results for the correction factor $f = {\cal N}/{\cal N}_{\infty}$. Dashed line: $f_V(X)$ derived from direct numerical simulations in Ref.~\protect \cite{Ve04}. Solid line: $f(X)$ (see Eq.~\ref{eq:f}) with $a = 0.275$ and $b = 0.39$ derived from the present experiments with the large apparatus. Dash-dotted lines: $f(X)$ (see Eq.~\ref{eq:f}) derived from the present experiments with the medium apparatus (upper curve: $\Gamma = 1.00$, lower curve: $\Gamma = 0.28$).}\label{fig:f_of_X}

\caption{The compensated Nusselt number ${\cal N} /R^{0.3}$ for the $D = 49.7$ cm diameter sample on a linear scale as a function of the Rayeigh number $R$ on a logarithmic scale. Open symbols: measured values of ${\cal N}$. Solid symbols: estimates of ${\cal N_\infty}$ obtained by correcting for the effect of the top and bottom plates on the heat transport in the fluid. Circles: copper top and bottom plates. Squares: aluminum top and bottom plates. In (a)  the formula suggested in Ref.~\protect \cite{Ve04} (Eq.~\ref{eq:f_V}, dashed line in Fig.~\ref{fig:f_of_X}) was used. In (b) Eq.~\ref{eq:f} with $a = 0.275$ and $b = 0.39$ (solid line in Fig.~\ref{fig:f_of_X}) was used. From left to right, the data sets are for $\Gamma = 3.0,$ 1.0, and 0.43. For comparison we show  the prediction of Ref. \protect \cite{GL01} based on a fit to the data from Ref.~\protect \cite{AX01} as solid lines.}\label{fig:comp1}

\end{figure}

\begin{figure}

\caption{The compensated Nusselt number ${\cal N} /R^{0.3}$ for the $D = 24.8$ cm diameter sample on a linear scale as a function of the Rayeigh number $R$ on a logarithmic scale. From left to right, the two data sets are for $\Gamma = 1.00$ and 0.275. Open symbols: measured values of ${\cal N}$. Solid symbols: estimates of ${\cal N_\infty}$ obtained by correcting for the effect of the top and bottom plates on the heat transport in the fluid. Circles: copper top and bottom plates. Squares: aluminum top and bottom plates. Equation~\ref{eq:f} with $a = 0.378~(0.304)$ and $b = 0.488~(0.506)$ (dash-dotted lines in Fig.~\ref{fig:f_of_X}) was used for $\Gamma = 1.00~(0.275)$.}\label{fig:comp2}

\caption{The measured (i.e. uncorrected) compensated Nusselt numbers ${\cal N} /R^{0.3}$ for the $D = 9.21$ cm diameter sample ($\Gamma = 0.967$) on a linear scale as a function of the Rayeigh number $R$ on a logarithmic scale. Circles: copper top and bottom plates. Squares: aluminum top and bottom plates. }\label{fig:comp3}

\caption{Reduced Nusselt numbers ${\cal N}/R^{0.3}$ obtained in the small apparatus with acetone at 32$^\circ$C for $\Gamma = 1.00$. Open circles: data from Ref.~\cite{Ah01} with an aluminum bottom and a sapphire top plate (the data were re-analyzed with small changes of the series resistance and differ slightly from those originally reported). Open squares: new data obtained with an aluminum bottom and a sapphire top plate. Solid circles: new data obtained with copper top and bottom plates.  The conduction of the empty sample cell was subtracted, but no further correction for the effect of the side walls on the heat current was made. The small vertical bar represents one percent.}\label{fig:acetone}
\end{figure}

\newpage

\begin{figure}
 \includegraphics[width=15cm]{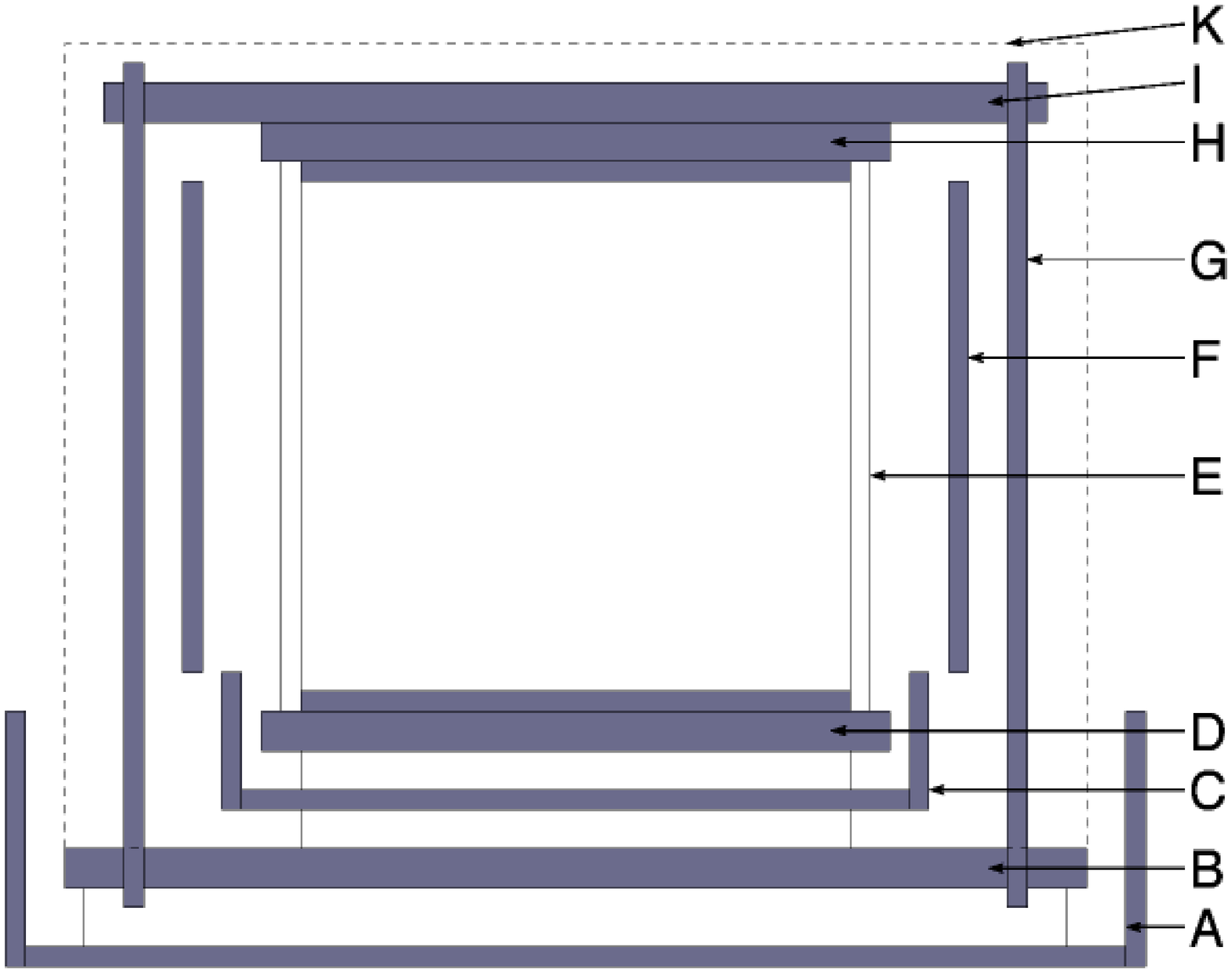}
 \centerline{Fig. 1}
 \vskip 4in
\end{figure}

\begin{figure}
 \includegraphics[width=15cm]{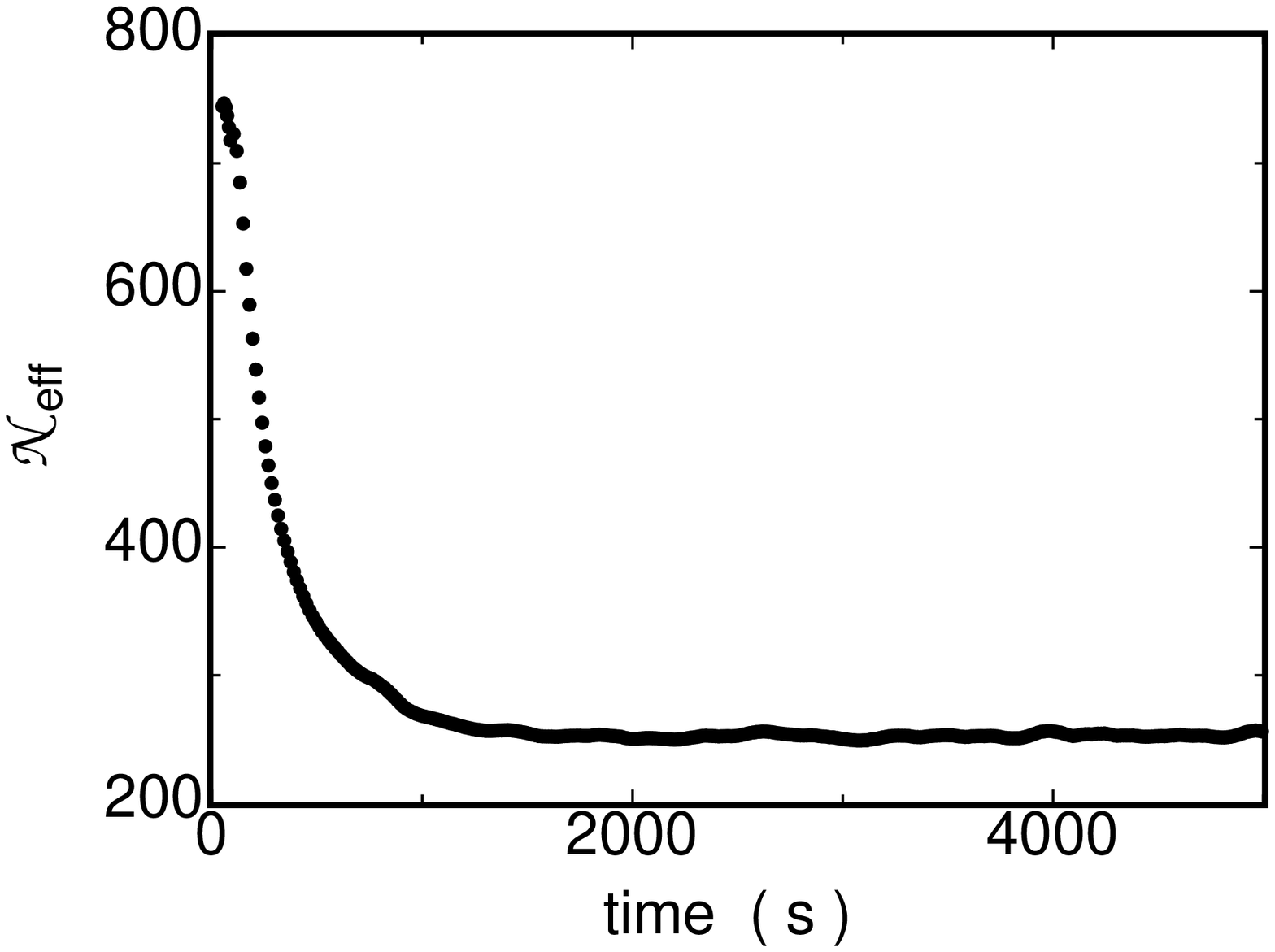}
 \centerline{Fig. 2}
\end{figure}

\begin{figure}
 \includegraphics[width=15cm]{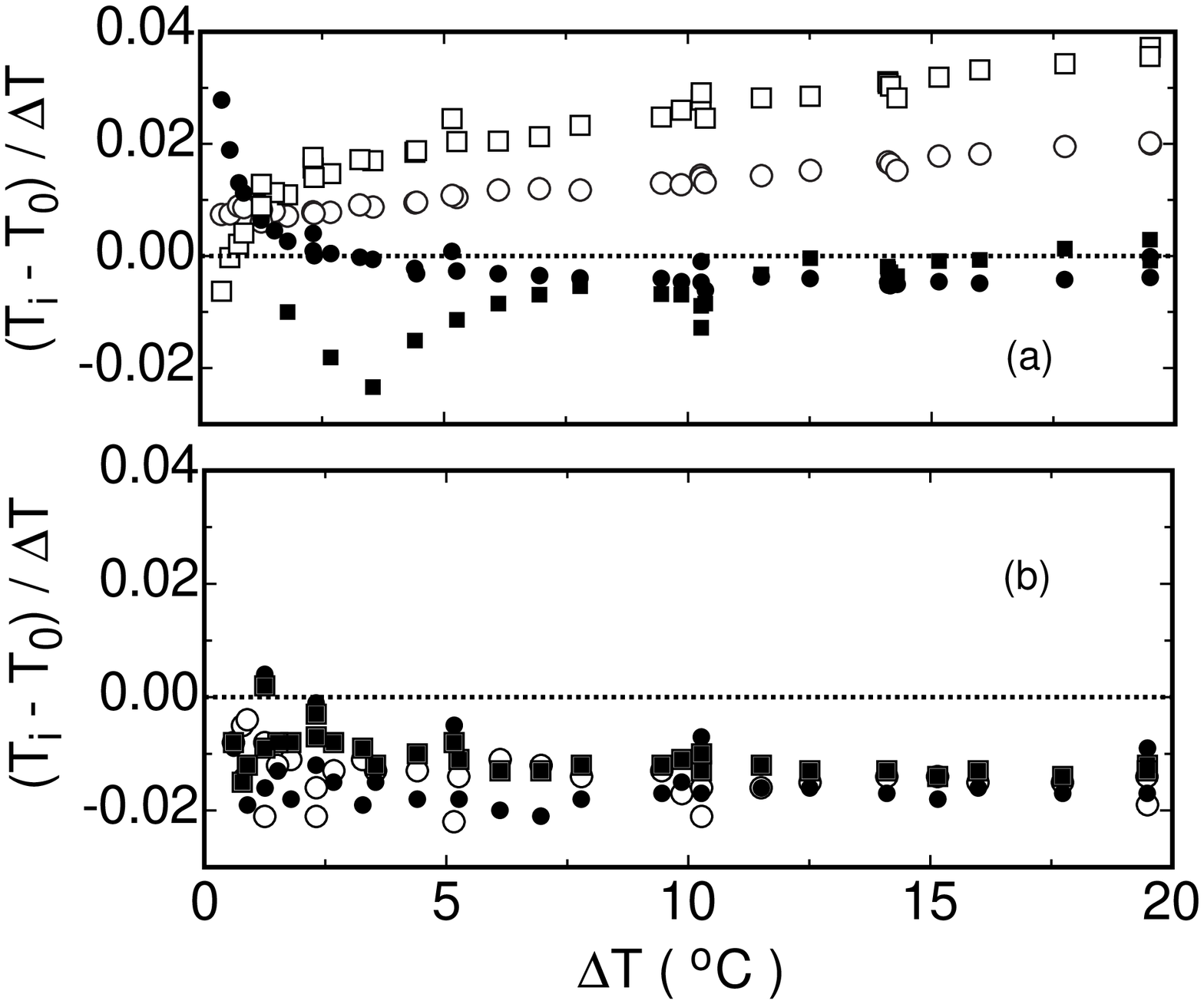}
 \centerline{Fig. 3}
\end{figure}

\begin{figure}
 \includegraphics[width=15cm]{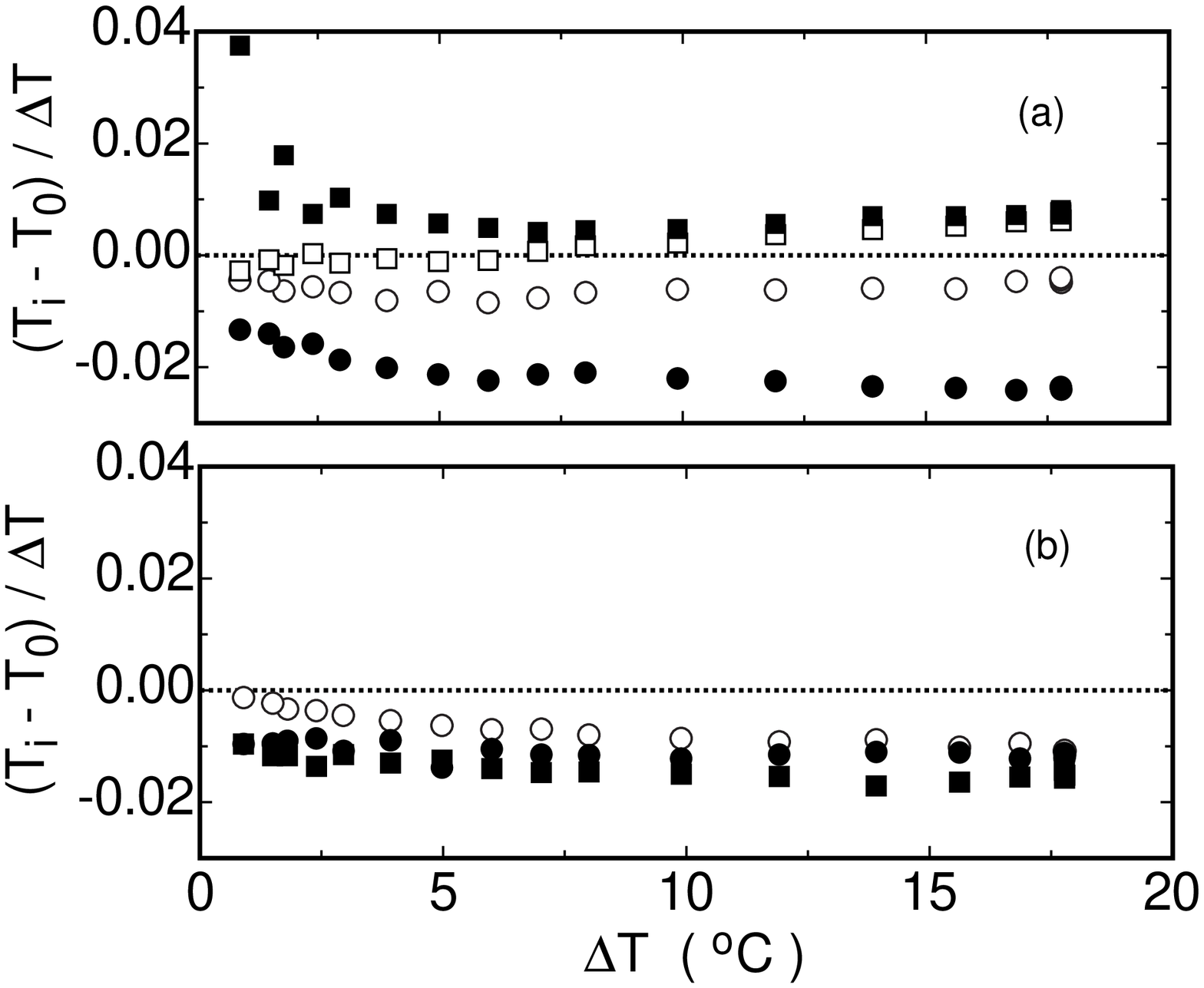}
 \centerline{Fig. 4}
\end{figure}

\begin{figure}
 \includegraphics[width=15cm]{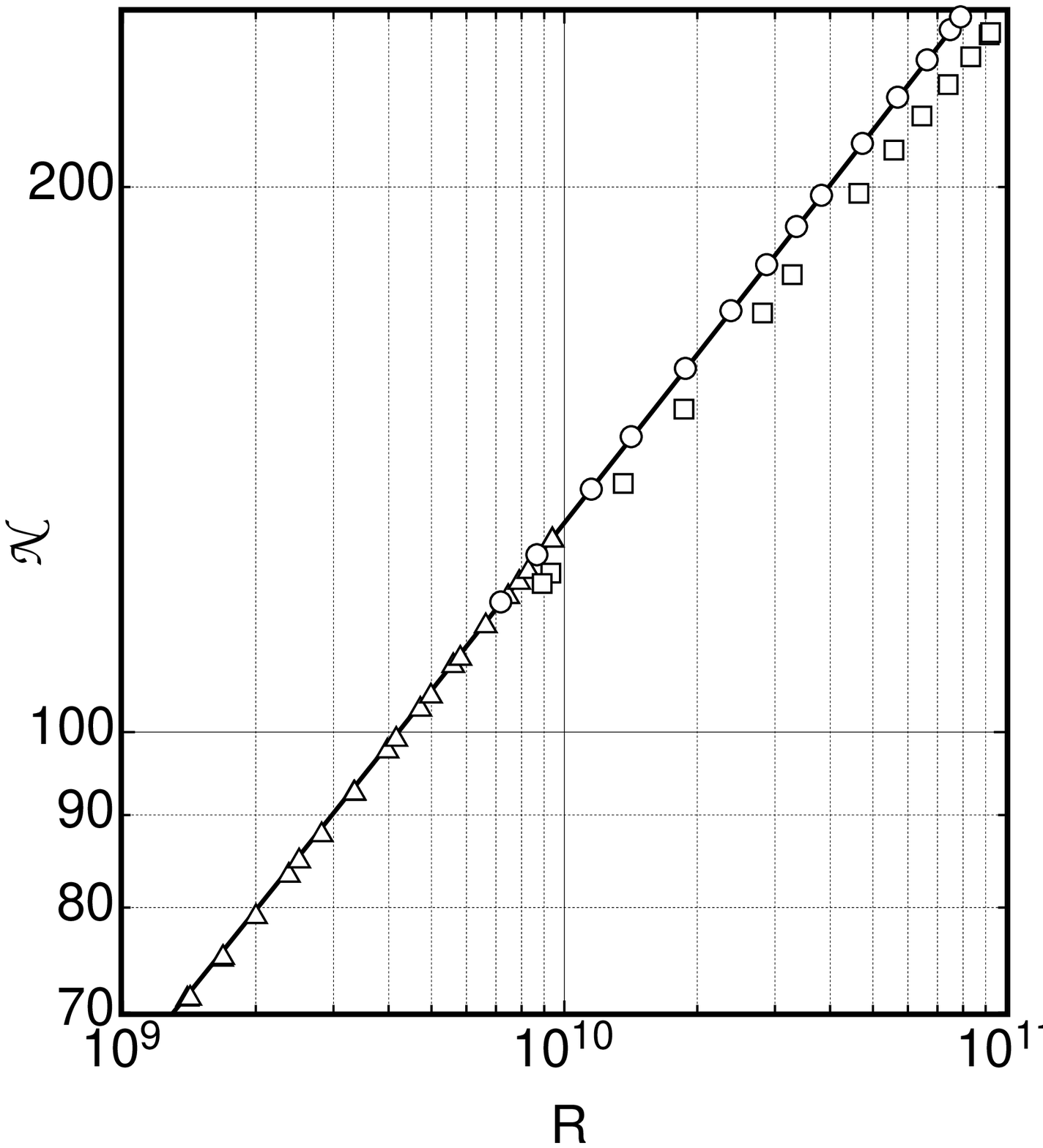}
 \centerline{Fig. 5}
\end{figure}

\begin{figure}
 \includegraphics[width=15cm]{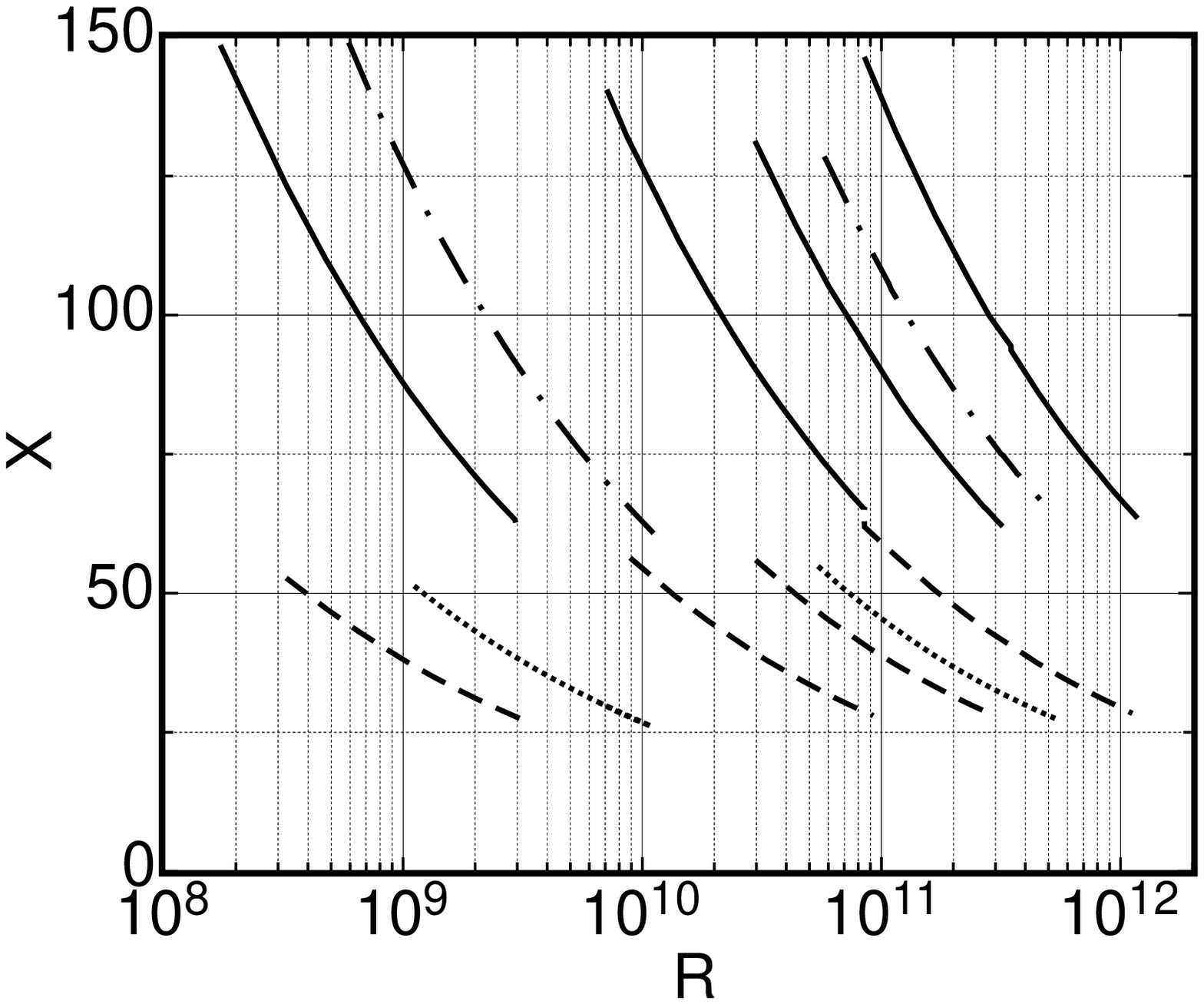}
 \centerline{Fig. 6}
\end{figure}

\begin{figure}
 \includegraphics[width=15cm]{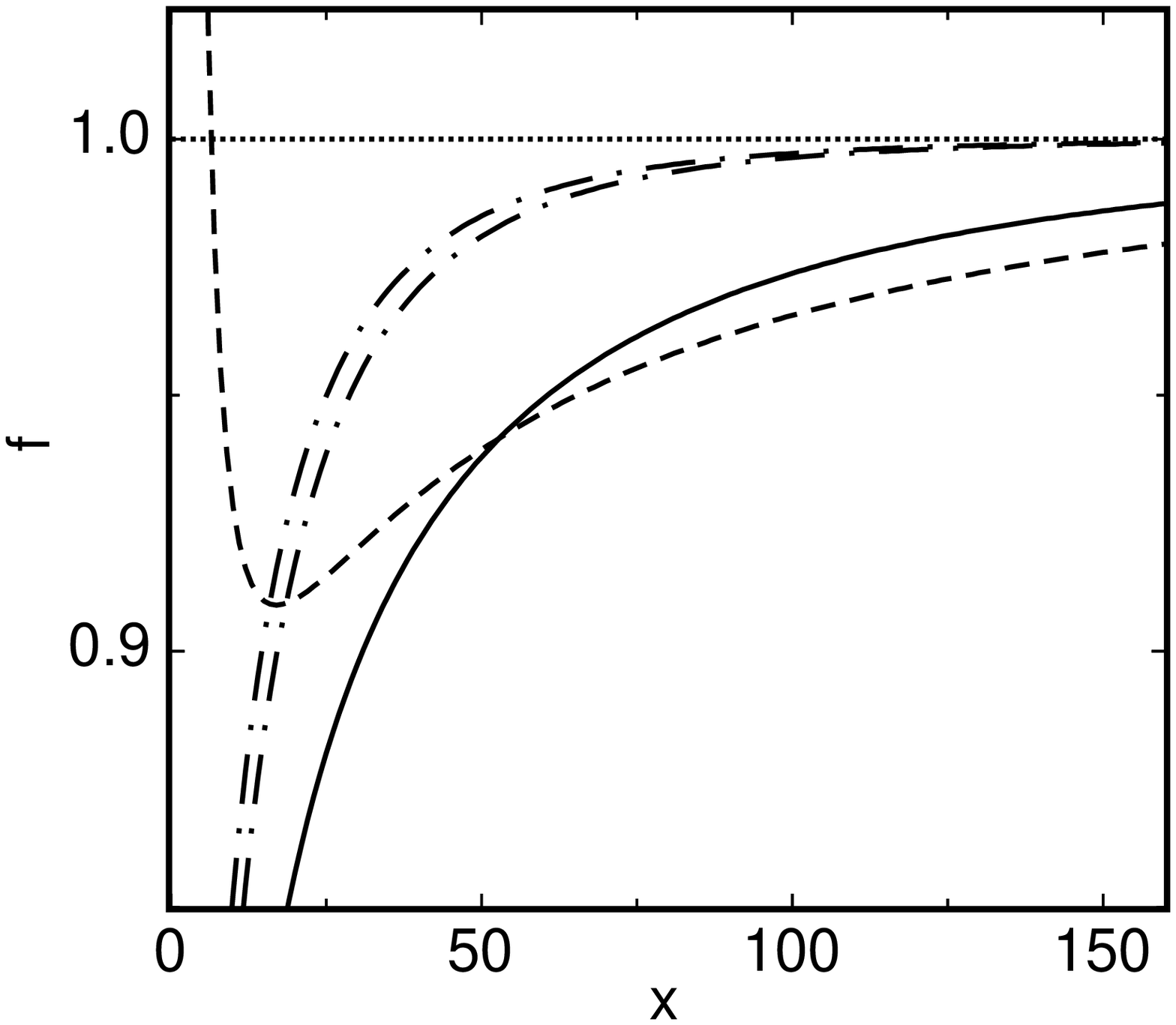}
 \centerline{Fig. 7}
\end{figure}

\begin{figure}
 \includegraphics[width=15cm]{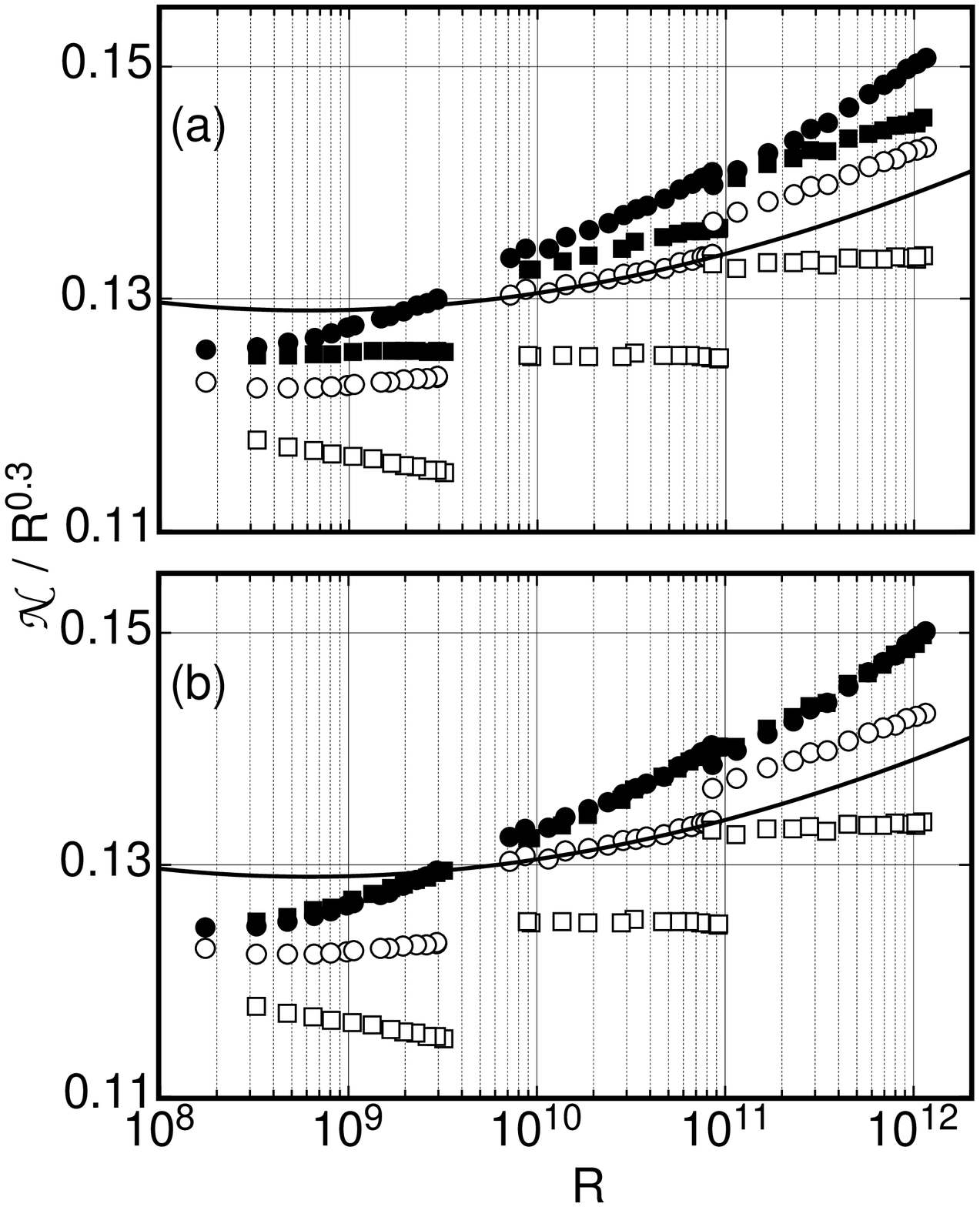}
 \centerline{Fig. 8}
\end{figure}

\begin{figure}
 \includegraphics[width=15cm]{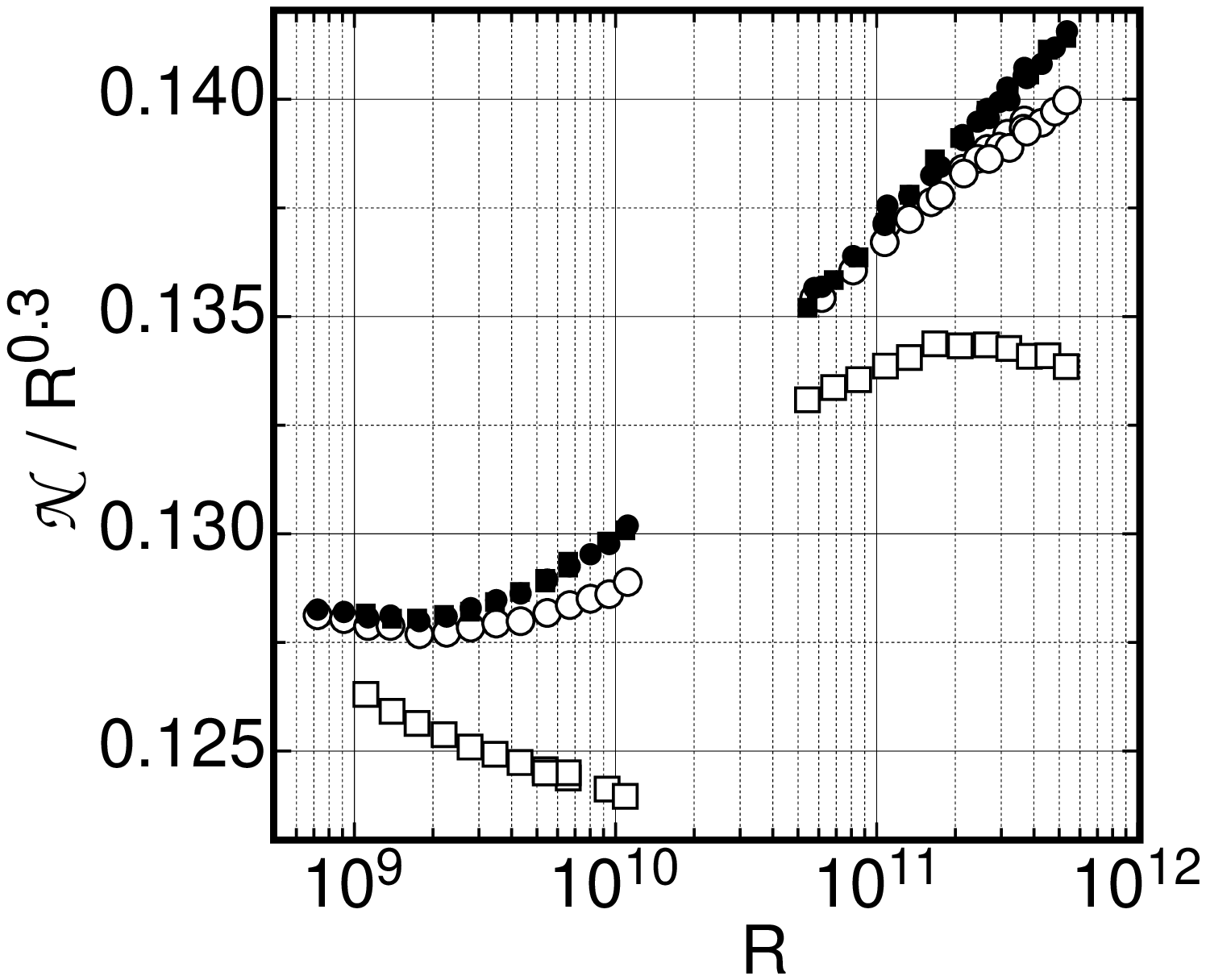}
 \centerline{Fig. 9}
\vskip 4in
\end{figure}

\begin{figure}
 \includegraphics[width=15cm]{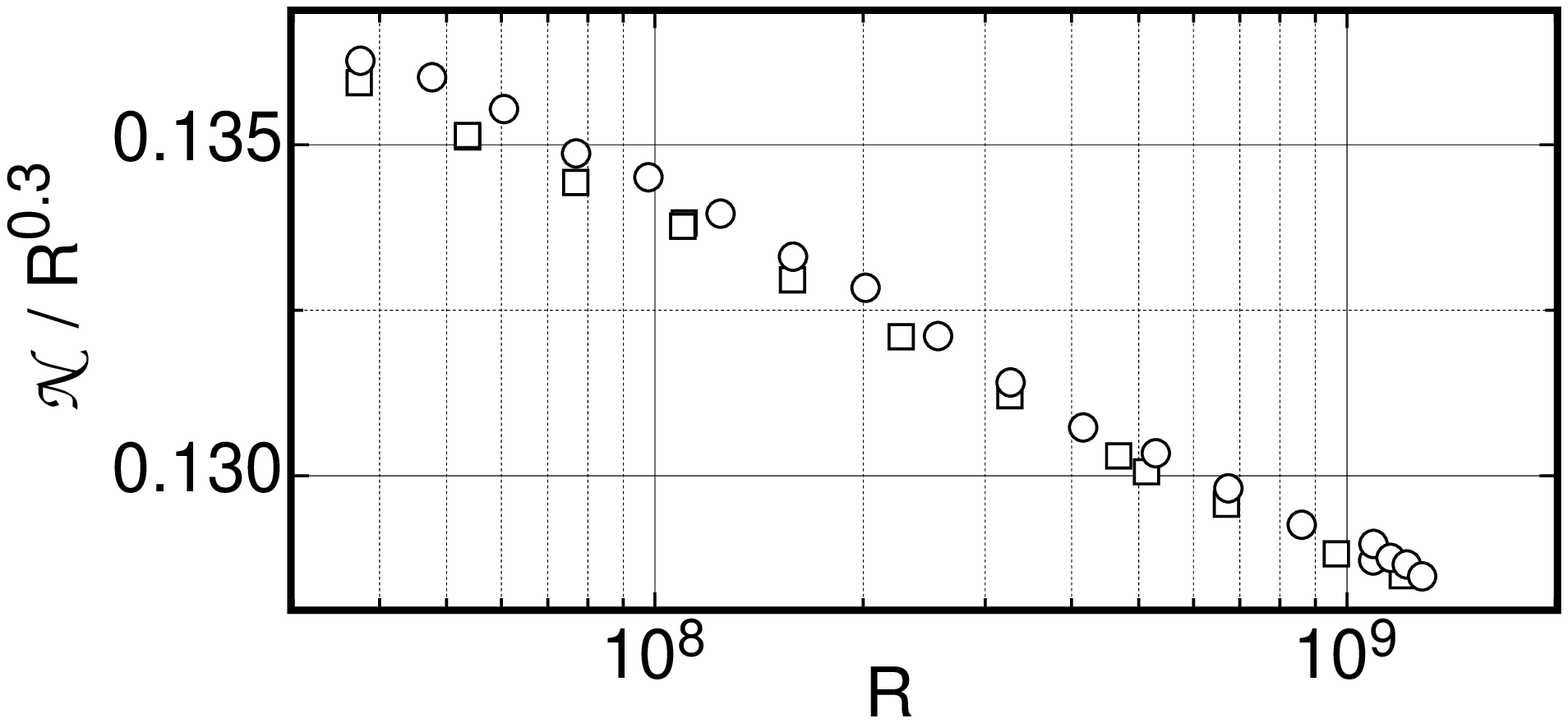}
 \centerline{Fig. 10}
\vskip 4in
\end{figure}

\begin{figure}
 \includegraphics[width=15cm]{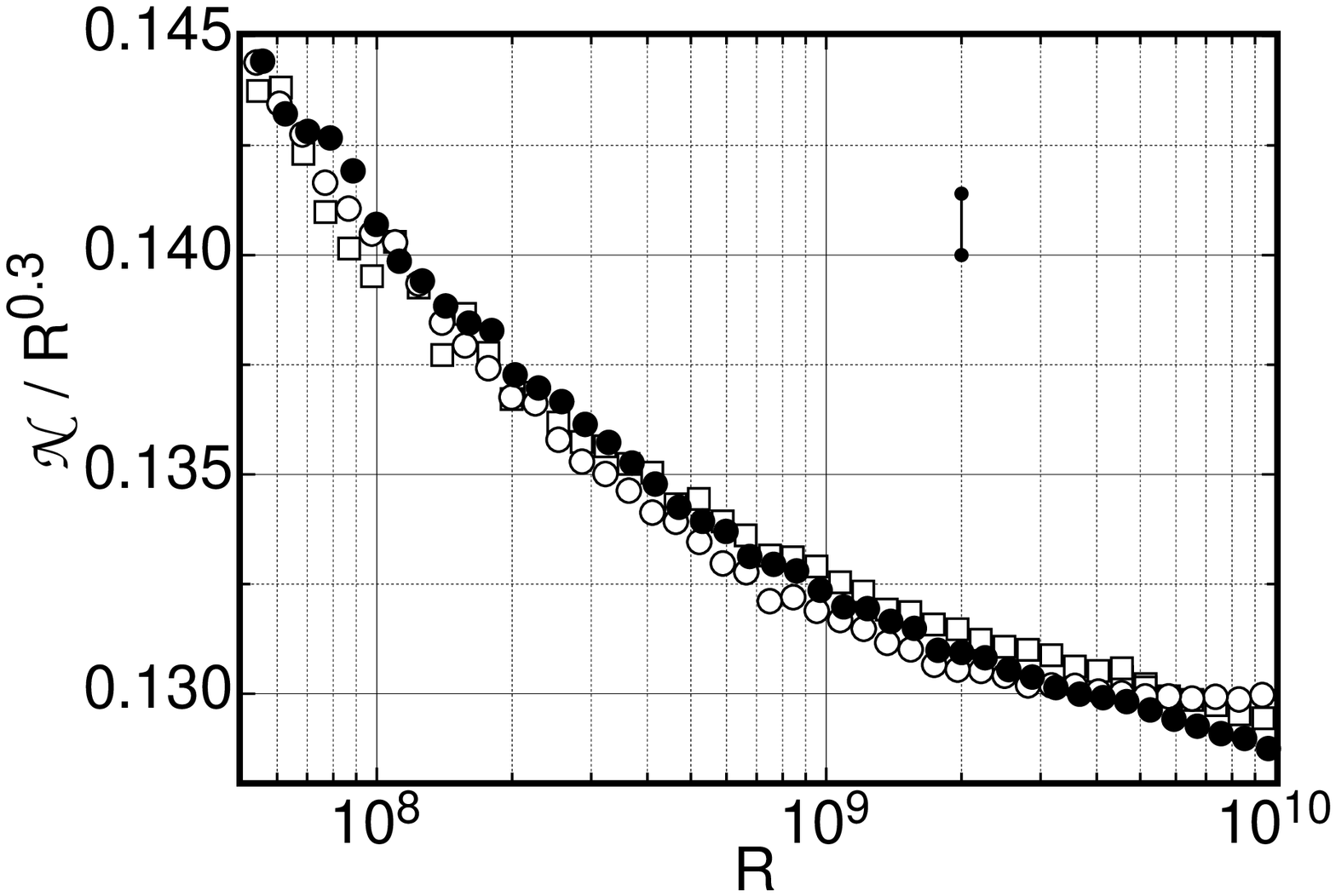}
 \centerline{Fig. 11}
\end{figure}


\begin{references}
\bibitem{med_htr} For the medium apparatus a circular area of diameter 24.8 cm was covered by parallel heater grooves of width 0.30 cm and depth 0.78 cm, spaced 0.51 cm apart. The heater wire was teflon-covered AWG18 Nichrome C and had a rsistance of 12.8 $\Omega$.
\bibitem{med_spiral} For the medium apparatus the channel width and depth were 0.95 cm and 2.54 cm respectively, and the spacing between adjacent arms was 1.9 cm.
\bibitem{Fenwal}We used Fenwal 120-503JAJ-Q01 thermistors with a nominal resistance of 50 k$\Omega$ at 25$^\circ$C.
\bibitem{FNBL} The total thermal resistance of the bottom plate of area 60 cm$^2$, pressed against the sapphire window with a very thin layer of glycerol in between, was measured to be 0.0286 K/W. The bottom-plate and sapphire contributions were estimated to be 0.003 and 0.014 K/W respectively.  The glycerol layer was estimated to have a thickness of 5 $\mu$m, and in that case would contribute about 0.003 K/W (if glycerol had not been used and air had been between the plates, the air resistance would have dominated). This leaves about 0.009 K/W for the thermal resistance in the water bath between the top of the sapphire plate and the thermometer. The uncertainty of this result is quite large due mostly to the uncertainty of the thickness of the glycerol layer. The original data of Ref.~\cite{Ah01} were re-analyzed with this correction. The results differ slightly from those originally reported.
\end{references}
\end{document}